\documentclass[aps,letterpaper,10pt,twocolumn,showpacs,superscriptaddress,nofootinbib]{revtex4-1}
\usepackage{amsfonts,amssymb,amsmath,amsthm,graphicx}
\usepackage{url}
\usepackage{dsfont}
\usepackage{bbm}
\usepackage[usenames,dvipsnames]{xcolor}
\usepackage{nicefrac}
\usepackage{natbib}
\usepackage{float}
\usepackage{setspace}
\usepackage{multirow}
\usepackage{subfigure}
\usepackage[export]{adjustbox}
\usepackage{comment}
\usepackage[normalem]{ulem} 
\usepackage{todonotes}

\def\ket #1{\vert #1\rangle}

\newcommand{\x}{{\bf x}}
\newcommand{\n}{{\bf n}}
\newcommand{\s}{{\bf s}}

\def\virgolette #1{``#1"}

\newcommand{\beq}{\begin{equation}}
\newcommand{\eeq}{\end{equation}}

\begin{document}

\title{Enhanced security for multi-detector Quantum Random Number Generators}

\author{Davide G. Marangon}
\affiliation{Dipartimento di Ingegneria dell'Informazione, Universit\`a degli Studi di Padova, Padova, Italia}
\author{Giuseppe Vallone}
\affiliation{Dipartimento di Ingegneria dell'Informazione, Universit\`a degli Studi di Padova, Padova, Italia}
\affiliation{Istituto di Fotonica e Nanotecnologie, CNR, Padova, Italia}
\author{Ugo Zanforlin}
\affiliation{Dipartimento di Ingegneria dell'Informazione, Universit\`a degli Studi di Padova, Padova, Italia}
\author{Paolo Villoresi}
\affiliation{Dipartimento di Ingegneria dell'Informazione, Universit\`a degli Studi di Padova, Padova, Italia}
\affiliation{Istituto di Fotonica e Nanotecnologie, CNR, Padova, Italia}

\begin{abstract}
Quantum random number generators (QRNG)
represent an advanced solution for randomness generation,
essential in every cryptographic applications.
In this context, integrated arrays of single photon detectors have 
promising applications as QRNGs based on the spatial detection of photons. 
For the employment of QRNGs in Cryptography, it is necessary to 
have efficient methods to evaluate the so called quantum min-entropy that corresponds to
the amount of the true extractable quantum randomness from the QRNG.
Here we present an efficient method that allow 
to estimate the quantum min-entropy for a multi-detector QRNG.
In particular, we will consider a scenario in which
an attacker can control the efficiency of the detectors and
knows the emitted number of photons. 
Eventually, we apply the method to a QRNG with $10^3$ detectors.
\end{abstract}
%

\maketitle

\section{Introduction}
Randomness is the fundamental ingredient of any cryptographic protocol.
Classical, quantum and even post-quantum algorithms are indeed based on the availability of genuine and secret random numbers. 
Cryptographic random number generators are required to meet two main requisites: the first is to generate unbiased and uncorrelated random numbers; the second requirement is to  provide no information about its output to the \virgolette{environment}. 
In this respect, random number generators based on quantum physical processes (QRNG) are considered more secure than chaos-based RNGs and algorithmic-RNGs: indeed, for QRNGs, unpredictability comes from the intrinsic probabilistic nature of a measurement in Quantum Mechanics.
However, realistic implementations of QRNGs 
must take into account \virgolette{imperfections} of the physical devices used in the generator. 
The latters might introduce statistical flaws, e.g. bias and correlations in the generated numbers and, more importantly, might provide side information to an attacker.
Hence, it is of paramount importance to estimate 
the leakage of information of the QRNG (also called side information)
in order to discard the bits which might have been guessed by an eavesdropper.

The purpose of this work is the evaluation of such side information and the extraction of random numbers for a QRNG based 
on an integrated array of single photon detectors.
In the framework of discrete variable QRNG, this scheme is promising because it is equivalent to \virgolette{parallelize} multiple generators \cite{stuc13spie} and indeed it has been explored 
 in recent 
works~\cite{burr13iisw,yan14rsi,sang14prx,li16cpl}.
However, an advanced security model for the estimation of the maximal extractable randomness, was still missing. 
In this work we provide it, by generalizing a novel paradigm recently introduced in \cite{frau13qph} for the side information of a two-detector QRNG. 
We will present a computationally efficient method  to estimate the quantum min-entropy for systems with a generic number of detectors. 
We finally experimentally applied our method to a \emph{single photon camera} that we converted into a QRNG. 

The paper is organized as follows: in Section \ref{Sec_1} we describe how \emph{raw} random numbers can be generated with the single photon camera system. 
In Section \ref{Sec_2.2} we introduce the enhanced security model for the QRNG, which is based on the estimation of its  min-entropy conditioned on the \emph{accessible} side information. 
In Section \ref{Sec_3}, we show how the entropy can be efficiently evaluated by a combinatorial approach.
In Section \ref{Sec_4} the new paradigm is applied to the QRNG and the results are discussed.

\section{Methods}
\subsection{Raw randomness generation} \label{Sec_1}
Our generator is based on the detection of photons by an array of single photon detectors. 
The system consists of a light source which uniformly illuminates 
the \emph{Single Photon Counting Camera} SPC$^2$ manufactured by Micro Photon Devices~\cite{guer10spie,SPC2}
and originally intended for the acquisition of images at low level of illumination.
Random numbers are generated according to which detector has clicked during a given time interval, 
as detailed below. 
The Camera is based on $M = 1024$ single photon detectors arranged in a matrix of 
$32\times32$ pixels. Every pixel
is equipped with a single photon avalanche detector (SPAD)
and its quenching circuit. 
Contrarily to normal CMOS or CCD sensors, where many photoelectrons are accumulated and converted in a digital value, here every single detected photon increases a counter associated to every pixel.

\begin{figure*}[!htb]
\centering\includegraphics[width=1.6\columnwidth]{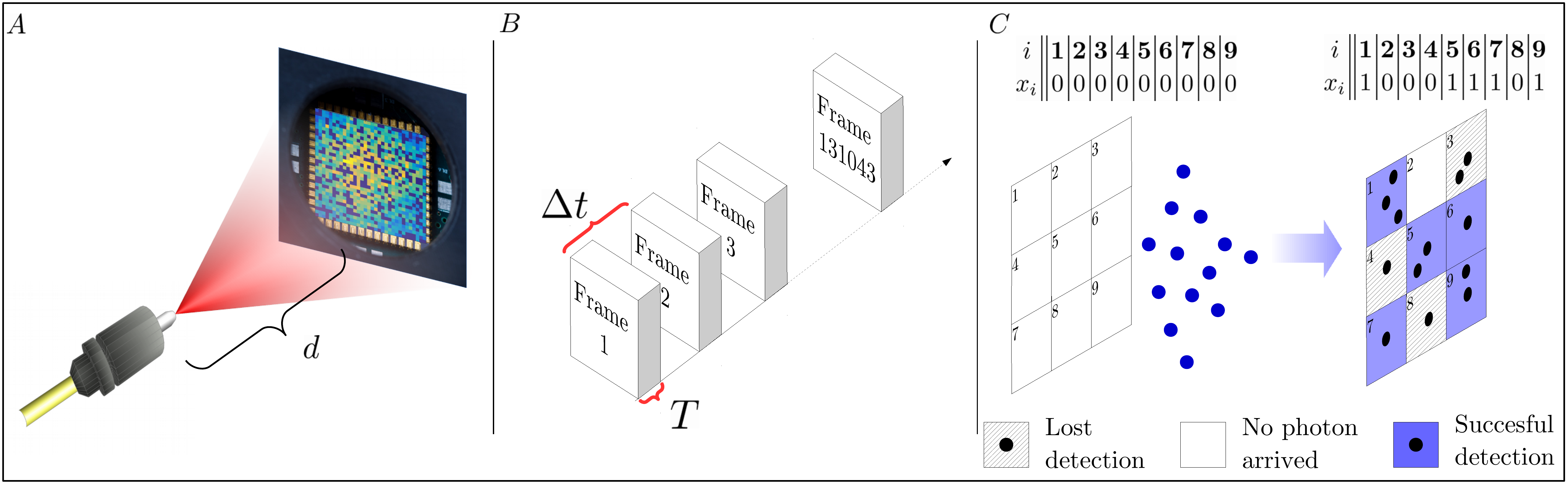}
\caption{Box A: Illustration of the SPC2 camera illuminated by the 808 nm light coming from a single mode fiber. 
Box B: the temporal structure of an acquisition is illustrated. Box C: The process of random number generation is reported for a single frame.}
\label{figura_1}
\end{figure*}

Raw random numbers are  generated by using the setup schematically reported in Fig.\ref{figura_1}-A. 
A laser ($\lambda = 808$ nm) coupled with a SM-optical fiber, is arranged in front of the camera to provide a controlled photon flux on the sensor. 
The laser intensity is not stricktly uniform on the array as it features the typical Gaussian profile of a TEM00 mode.
However, according to the distance $d$ between the fiber output and the sensor, the laser intensity can be considered \virgolette{locally} uniform on $M^\prime<M$ pixels (where $M^\prime$ increases as function of $d$; the data considered here were collected for $d=10$ cm).
The SPC$^2$ detects the light by acquiring multiple \virgolette{frames} as we illustrate in Fig.\ref{figura_1}-B.
Every frame is characterized by an integration interval $T$, 
that we set to $T = 200$ ns (the shortest integration interval achievable by the camera).
Frames are obtained at a rate of $R_{\text{acq}}= 49$ kHz, corresponding to a 
temporal distance between two frames of $\Delta t=1/R_{\text{acq}}$.
The counter of each pixel detector $D_i$ in each frame can take only two values, $x_i=0$ and $x_i=1$, corresponding to 
a non-detection and a photon-detection respectively.
Hence, for each frame, a random number  $\x=\{x_1,\dots,x_M \}$ is generated by concatenating the pixel outputs in a string $M$ bits long.
We indicate by X the random variable that takes the $\x$ values.
To illustrate the process, in Fig.\ref{figura_1}-C, we show
a sub-matrix of $3 \times 3$ pixels.
In the figure the sub-matrix is characterized by $n=9$ incoming photons from the laser which are independently acquired by the pixel detectors.
Due to inefficiencies of the detectors or dead-time, some of the incoming photon are not
registered by the pixel detectors.

The probabilities to obtain $x_i=0$ or $x_i=1$ from a pixel $D_i$ are respectively denoted by 
$\mathcal P^{(i)}_0$ and $\mathcal P^{(i)}_1$.
Such probabilities depend on the mean photon number $\mu_i$ at detector $D_i$ and from the detector features.
By properly adjusting the $\mu_i$ is then possible to obtain an \virgolette{unbiased} sequence, 
i.e. $\mathcal P^{(i)}_0=\mathcal P^{(i)}_1=1/2$. 
In order to simplify the evaluation of the probability $\mathcal P(\x)$ of generating a string $\x$, 
it can be assumed that on a small sub-matrix the light intensity is uniform, i.e. $\mu_i\approx \mu$.
For the case considered in the scheme, we are considering the pixel square corresponding to the centre of the matrix.
Furthermore, as the SPADs are approximately characterized by the same efficiency, it can be also assumed that $\eta_i\approx \eta$.
With these assumptions, as the pixels detect the photons independently, we have that
$\mathcal P^{(i)}_0\equiv \mathcal P_0$ and $\mathcal P^{(i)}_1\equiv \mathcal P_1$, $\forall i$.
Then, the probability $\mathcal P(\x)$ of generating a string $\x$ with $k$ bits 1 and $M-k$ bits 0
is  given by $\mathcal P(\x)= \mathcal P^{M-k}_0\mathcal P^{k}_1$.

\begin{figure}[tb]
\centering\includegraphics[width=1\columnwidth]{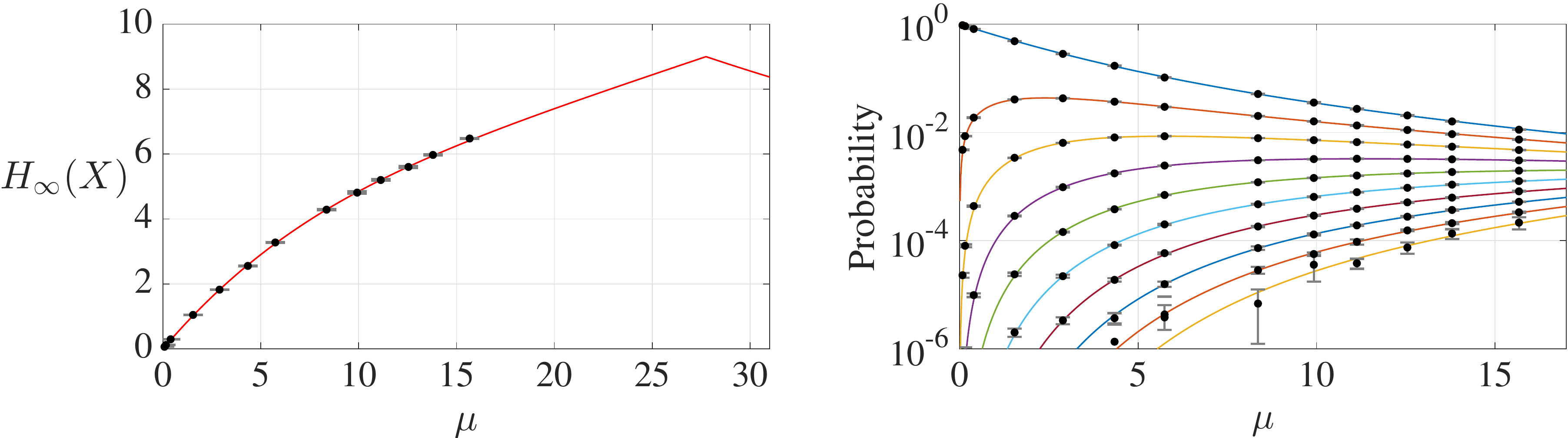}
\caption{(left) Min-entropy $H_{\infty}(X)$: experimental data (black dots) and fit (solid red line) are reported as a 
function of the mean photon number $\mu$ per pixel for $M=9$ pixels. (right) The $M+1$ inequivalent 
$\mathcal P(\x)=\mathcal P_0^{M-k}\mathcal P^k_1$ 
are reported for 
$k\in \left[0,9\right]$: the $\x$ corresponding to smaller $k$ are more likely with respect to the ones corresponding to larger $k$.
Indeed, the top curve corresponds to $k=0$ while the bottom one to $k=9$.}
\label{figura_2}
\end{figure}

If we assume that the QRNG is a perfectly isolated system, the number of random bits that can be obtained in each frame by the generator is measured by the so called classical min-entropy
\beq
\begin{aligned}
H_{\infty}(X)&=- \log_2[ \max_{\x} \mathcal{P}(\x)]\,.
\end{aligned}
\eeq
When $\mathcal P(\x)= \mathcal P^{M-k}_0\mathcal P^{k}_1$, the classical min-entropy can be easily evaluated as
$H_{\infty}(X)=- M\log_2[ \max\{\mathcal P_0, \mathcal P_1\}]$.
The unbiased randomness rate thus becomes  $R_{\text{gen}}=H_{\infty}(X) R_{\text{acq}}$. 
In Fig.\ref{figura_2} (left) we report $H_{\infty}(X)$ as a function of $\mu$  with the relative experimental data. 
In Fig.\ref{figura_2} (right), the ten inequivalent probabilities $\mathcal P(\x)$, for $k\in \left[0,9\right]$ are plotted: 
for the  values of $\mu$  (from 0 to 15 photons/pixel) considered in the experiment, the most likely outcome
is the string with all bits equal to zero. 
The accordance of the experimental points with the theoretical prediction shows that the 
approximations $\mu_i\approx \mu$ and $\eta_i\approx \eta$ are well justified.
To achieve the maximal rate, a proper value of $\mu=\mu_{\text{opt}}$ should be used, which enables an uniform distribution of the outcomes $\x$. 
For the case considered in Fig.\ref{figura_2}, such value is $\mu_{\text{opt}}\approx 28$ photons per pixels.
For this optimal value it is possible to achieve the maximum generation rate of $R_{\text{gen}}=M R_{\text{acq}}=50.176~Mbit/s$ 
corresponding to a min-entropy $H_{\infty}(X)=M$.
However, when the QRNG is used in a cryptographic scenario, the content of true random bits must be evaluated
in a different way and the maximum
generation rate is achieved with lower values of $\mu$, as demonstrated below.

\begin{figure*}[t]
\centering\includegraphics[width=0.8\textwidth,frame]{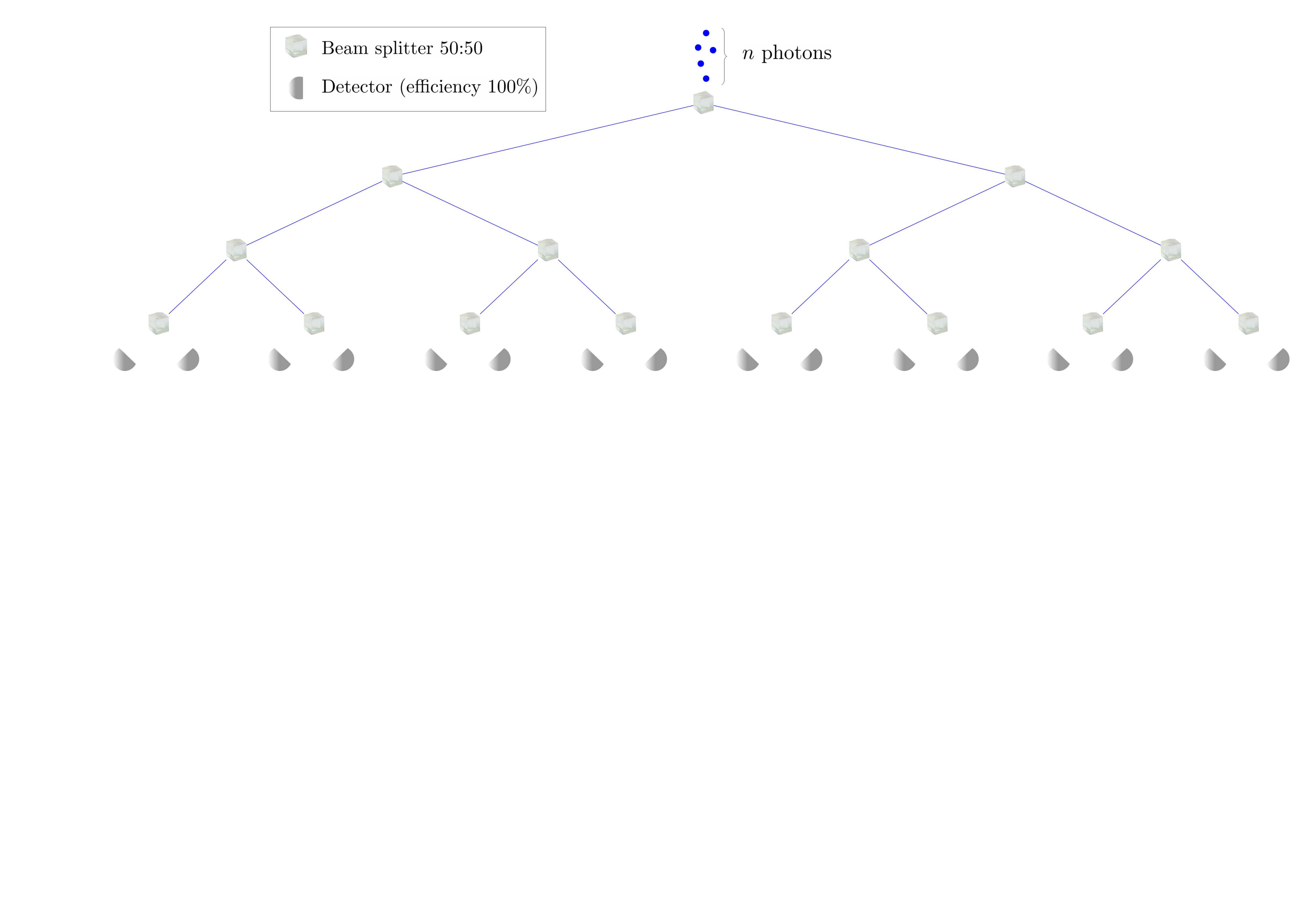}
\caption{Equivalent optical model of our QRNG, based on a photonic quantum state characterized by an uniform coherent spatial modal structure detected by a matrix of pixels. The equivalent model is the 50:50 beam splitter \virgolette{multinomial tree}, whose scheme is reported in Figure.}
\label{figura_3}
\end{figure*}

\subsection{Secure randomness generation}
\label{Sec_2.2}
For cryptographic applications, 
the classical min-entropy $H_{\infty}(X)$ cannot be used and more advanced measures of randomness are needed:
indeed, beside the requirement of being independent and identically distributed, it is also necessary that the numbers are not known to anyone else but the legitimate user.
In this case, it is necessary to evaluate the so called conditional quantum min-entropy $H_{\text{min}}(X|E)$ 
\cite{koni09ieee,toma10ieee,ren12ieee}.
The obtained random sequence will be post-processed with a seeded randomness extractor,
obtaining a random stream whose lenght is determined by $H_{\text{min}}(X|E)$.

The quantum min-entropy is related to the correlations that the quantum system has with the environment $E$ that
can be controlled by an adversary.
These possible correlations follow from non-idealities of the QRNG devices: for instance the input state may be mixed 
and the performed measurements may correspond to positive operator value measurements (POVM) rather than projectors.
For ideal QRNGs, where the random numbers are generated by measuring an isolated system with devices implementing
perfect projector operators, the 
$H_{\text{min}}(X|E)$ reduces to the classical min-entropy. 
However, for realistic QRNGs, $H_{\text{min}}(X|E)$ cannot be exactly known and a typical approach consists in finding a lower bound. 
For instance, by assuming trusted measurement devices and a completely untrusted source, a lower bound on $H_{\text{min}}(X|E)$ can be found \cite{vall14pra2, mara15qph}.

In \cite{frau13qph}, Frauchiger {\it et al.} addressed this problem for the \virgolette{welcher weg} QRNG, implemented by a diagonally polarized photon impinging on a polarizing beam splitter (PBS): depending on its polarization (horizontal $H$ or vertical $V$), the photon is transmitted or reflected by the PBS.
There, the photonic state is assumed to be pure but the measurements on the PBS output spatial modes are described by POVMs.
In this framework, the residual randomness is measured by means of $H_{\text{min}}(X|C)$.
The classical random variable $C$ encodes \emph{classical information} about the degrees of freedom associated to the hardware non-idealities.
In particular, $C$ describes the state of the generator with respect to
the multi-photon emission from the light source, (i.e. the number of emitted photons)
and the detectors inefficiencies.
The significant advantage of this paradigm is that $C$ is accessible and measurable by the QRNG user. 
Furthermore, the authors demonstrate that $H_{\text{min}}(X|E)\geq H_{\text{min}}(X|C) $, i.e. by conditioning on the classical information the QRNG is at least as secure as by conditioning on $E$.
In the following we will  show that such model can be adapted to the camera QRNG.

The model that we are considering for the multi-pixel QRNG is thus the following:
\begin{itemize}
\item the eavesdropper has information on the photon number emission of the source; 
\item the eavesdropper may determine the activation of single pixels during each acquisition;
\end{itemize}

We now explain in detail the above mentioned model. The multi-photon emission is characterized by the probability distribution $\mathcal{P}_N(n)$ of the classical random variable
$N\in\mathbb N$, which depends on the source itself (for a laser source $\mathcal{P}_N(n)=e^{-\mu}\mu^n/n!$). 
Eve is supposed to know the number of photons $n$ emitted in each pulse.
Moreover, detectors inefficiencies and dead times can be modelled by perfect detectors that are activated, with a given probability $\widetilde\eta_i$, 
by the eavesdropper.
In such scenario, the eavesdropper has the ability to deactivate any pixels $D_i$ in order to \virgolette{set} to 0 the bits $x_i$. 
The probabilities $\{\widetilde \eta_i\}$ must match 
the measured probabilities $\mathcal P^{(i)}_0$ and $\mathcal P^{(i)}_1$ of obtaining 0 or 1 
at pixel $i$, namely
\begin{align}
\label{prob_uno_Renner}
\mathcal P^{(i)}_0
&=1-\widetilde{\eta}_i(1-\mathbb P_0^{(i)})\,, 
\qquad\mathcal P^{(i)}_1
=\widetilde{\eta}_i(1-\mathbb P_0^{(i)})\,.  
\end{align}
In the above equations, $\mathbb P_0^{(i)}$ is the probability that no photon arrives at the pixel detector $i$.
The probabilities $\widetilde{\eta}_i$ can be regarded as an \virgolette{equivalent efficiency} which accounts for the effects of both the quantum efficiency of the SPAD, $\eta_i$, and its dead time.
\begin{figure}[h]
\centering\includegraphics[width=\columnwidth]{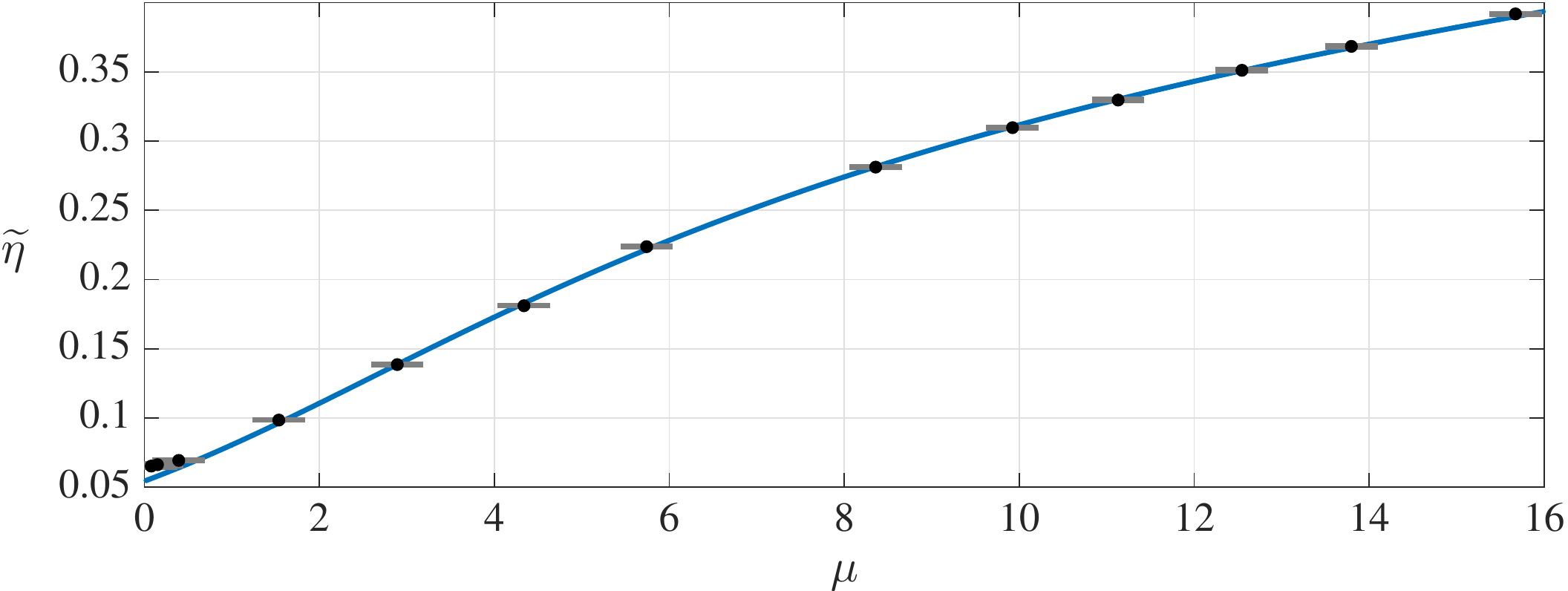}
\caption{
Values of $\widetilde{\eta}_i$ in function of $\mu$ as derived by  Eq. (\ref{prob_uno_Renner}).
The dots represent the experimental data, while the curve is the fit.
}
\label{figura_extra}
\end{figure}
Depending on $\mu$, the probability $\mathbb P_0^{(i)}$ will change and
Eve adjusts $\widetilde{\eta}_i$ in order to satisfy 
Eq. (\ref{prob_uno_Renner}).
This can be understood by considering Fig.\ref{figura_extra}, where the value of  $\widetilde{\eta}$ is reported as function of $\mu$.
We associate a random variable $s_i$ to the status of a pixel $D_i$, such that $s_i=0$ labels the inactive status, while 
$s_i=1$ corresponds to the active status. 
In each frame, Eve selects a pixel configuration vector 
$\s=\{s_1,s_2,\dots,s_M\}$ with probability
$\mathcal P_S({\bf s})=\prod_{i=1}^M\widetilde{\eta}_i^{s_i}(1-\widetilde{\eta}_i)^{1-s_i}$
from the set $S=\{ \s \}$ of cardinality $2^{M^{\vphantom{A}}}$.

Within this framework, the accessible classical information is therefore identified by $C=\{N,S\}$.
We will demonstrate
that, even if the eavesdropper has this information at hand, secure random numbers may be generated.
In our model, additional hardware defects, such as after-pulses (intended as the enhanced probability of having a detection at frame $j+1$
if the pixel clicked at frame $j$) and cross-talks among the pixels, are not included.
Indeed their effect can be neglected since we measured a negligible after-pulse probability and 
a probability $p_{\text{ct}}\approx  10^{-4}$ for the cross-talk.
By using the approach introduced in~\cite{frau13qph}
the min-entropy $H_{\text{min}}(X|C)$ can be evaluated by the following relation:
\begin{align}\label{eq_entropy}
2^{-H_\text{min}(X|C)}
=\sum_{n=0}^{\infty}\sum_{\{s_i\}}\mathcal{P}_N(n)
\mathcal{P}_{S}({\bf s})
2^{-H_\text{min}(X|n,{\bf s})}
\end{align}
where
\begin{align}
\label{H(X|n,s)}
2^{-H_\text{min}(X|n,{\bf s})}
=\max_{\bf x}\mathcal{P}(\x|n,{\bf s})\,,
\end{align}
and $\mathcal{P}(\x|n,{\bf s})$ is the probability of obtaining the random string $\x$ conditioned on the chosen
pixel configuration vector ${\bf s}$ and the emitted number of photons $n$.
In the above equation the sum $\sum_{\{s_i\}}$
runs over all the $2^M$ possible status configurations. 
We note that, if Eve has no information on the emitted number of photons, the min-entropy
should be evaluated as
$2^{-H_\text{min}(X|C)}
=\sum_{\{s_i\}}\mathcal{P}_{S}({\bf s})
2^{-H_\text{min}(X|{\bf s})}$ with $2^{-H_\text{min}(X|{\bf s})}
=\max_{\bf x}[\sum_{n=0}^{\infty}\mathcal{P}(\x|n,{\bf s})\mathcal{P}_N(n)]$.\phantom{bla bla bla b}
\linebreak

Eq. (\ref{eq_entropy}) generalizes the relation introduced in \cite{frau13qph} from two detectors to $M$ detectors.
From Eq. \eqref{eq_entropy}, the conditional entropy $H_\text{min}(X|C)$ depends on the largest conditional probability 
$\mathcal{P}(\x|n,\s)$. 
In fact, according to our model, 
knowing the number of photons emitted by the source and the sensor configuration,
the best guessing strategy for Eve is to bet on the string $\x$ with the largest probability of appearance. 
As described in Eq. (\ref{eq_entropy}), in order to obtain a lower bound on the conditional quantum min-entropy,
such maximal probability $\max_{\bf x}\mathcal{P}(\x|n,{\bf s})$
must be \virgolette{weighted} with the probability of having a given $\{n,\s\}$.
In the next section we will present our main result, which is an efficient method to evaluate the conditional probabilities 
$\mathcal{P}(\x|n,\s)$ (see in particular Eq. \eqref{eq_fact_rStirling}).

\section{Results}
\label{Sec_3}
In order to calculate $\mathcal{P}(\x|n,\s)$, we need to evaluate the evolution of $n$-photon Fock states and
how such photons arrive at the different detectors.
If the spatial wavefunction of each photon is pure,
the equivalent quantum optical model of our setup corresponds to the so-called \virgolette{multinomial tree} of 
$2^L-1$ beam splitters (with 50:50 reflection:transmission ratio) followed by $M=2^L$ detectors, 
as schematically reported in Fig. \ref{figura_3} for $L=4$\footnote{ If $M$ is not a power of 2, a multinomial tree can be also obtained. In this case the beam splitters are no more 50:50, but
their transmissivity is set such that a single photon entering in the tree has the same probability of arriving at each detector.}.  
Indeed, in the approximation of an uniform and coherent spatial structure of the light with respect to the sensor, 
a single photon emitted by the laser has the same probability of arriving at each pixel detector.
The requirement of the purity of the spatial wavefunction is essential for the assessment of the quantum randomness: only in this case
the detection of the photon in a given location is truly random and cannot (even in principle) be predicted.
On the other hand, if the spatial wavefunction is incoherent (such as it happens with a LED emission),
the information on the photon location is correlated with the environment and could, at least in principle, be predicted.
For this reason it is crucial that the light exits from a single mode fiber (the
mode propagating into the single-mode fiber is pure).
In line with \cite{frau13qph}, in the multinomial tree, the $n$ photons at the input evolves coherently towards the output
and nothing than vacuum states enters at the other inputs of the tree.

If $n$ photons are emitted by the source within the integration time $T$, the quantum state $\ket{\Psi}_n$ 
before the detectors can be written as
\begin{align}
\label{psi}
\ket{\Psi}_n &
=\frac{1}{\sqrt{n!}}\big(\frac{1}{\sqrt{M}}\sum_{i=1}^M\hat{a}^\dag_i\big)^n\ket{0}
\\
&=
\sum_{\{n_i\}}
\frac{\sqrt{n!} }{\sqrt{M^{n} n_1!\cdots n_M!}} \delta_{n_1+\ldots+n_M,n}
\ket{\n}\,,
\end{align}
where $\ket{0}$ is the joint vacuum of all modes, 
$\hat{a}_i^{\dagger}$ are the output modes creation operators and
 $\ket{\n}\equiv\ket{n_1,\cdots,n_M}$ is the normalized state with $n_i$ photons arriving at detector $D_i$.
We note that the Kronecker delta implements the condition $n=\sum_k n_k$.
In our model, the $M$ output mode entangled state $\ket{\Psi}_n$ is the equivalent of bi-mode entangled output state at the polarizing beam-splitter in \cite{frau13qph}.
In addition, because the tree has not common knots, interference effects are excluded; therefore the probability that all $n$ 
photons arrive at pixel $D_i$ is given by  $1/M^{n}$.

\begin{table}[t!]
\centering
\begin{tabular}{c|c||c}
 $x_i$ & $s_i$ & requirement on $n_i$ \\\hline
0 		& 0 		& no requirement \\\hline
0 		& 1 		& $n_i=0$ \\\hline
1 		& 0 		& no arrangement is compatible \\\hline
1 		& 1 		& $n_i\geq1$
\end{tabular}
\caption{The Table reports the requirement for each pixel detector $i$ for a {\it compatible} arrangement $\n$ given that
the output $\x$ is obtained and the detector configuration $\s$ is chosen.}
\label{tbl:requirements}
\end{table}

\begin{figure*}[t]
\centering
\includegraphics[height=5cm]{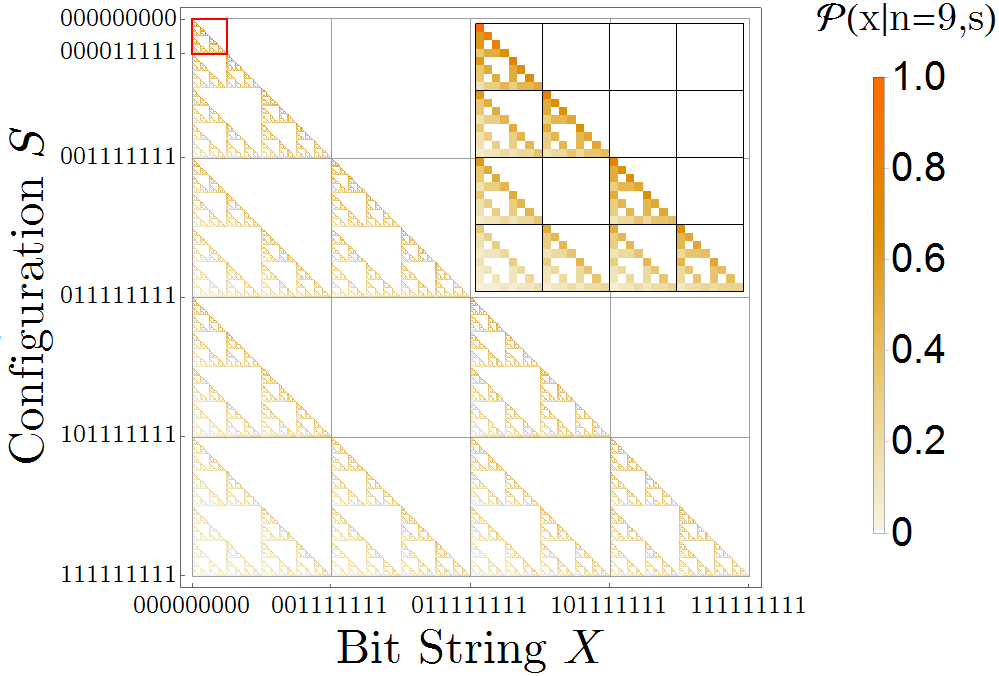}
\includegraphics[height=5cm]{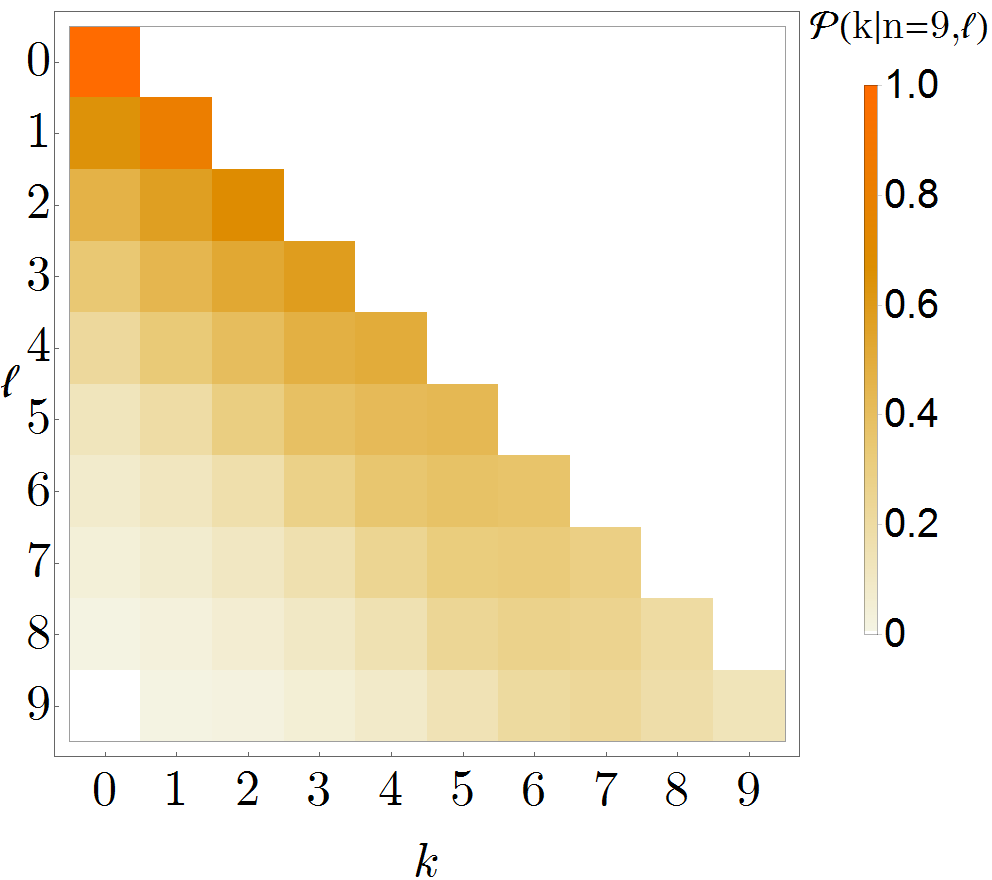}
\caption{\textbf{Left:} In a $512\times 512$ matrix, we show
the conditional probabilities $\mathcal{P}(\x|n,\s)$ for all the possible $\x$ (columns) and pixel status configurations $\s$ (rows), 
for a sensor matrix with $M=9$ and with $n=9$. The matrix resembles the multinomial Sierpinsky triangle. 
A zoom of the fractal structure is reported into the inset.
\textbf{Right:} Inequivalent values of $\mathcal{P}(\x|n,\s)$ are shown in function of
$k$ and $\ell$.}
\label{figura_4}
\end{figure*}

Because, as presented in \cite{frau13qph}, the detection operators applied on $\ket{\Psi}_n$ are all diagonal in the Fock basis, 
the estimation of the
probabilities $\mathcal{P}(\x|n,\s)$ 
can be approached classically by means of the multinomial distribution.
From Eq. \eqref{psi}, the probability that $n$ photons arrive at the different detectors 
with a given arrangement $\n=\{n_1,n_2,\cdots,n_M\}$ is indeed given by the multinomial distribution.
Then, for a given set of classical variables $\left\{n,\s\right\}$, $\mathcal{P}(\x|n,\s)$ can be computed by the sum of 
the probabilities of obtaining the arrangements $\n$ that are \emph{compatible} with the output $\x$ and the given detector configuration $\s$.
The term compatible means the following: if $x_i=1$, the \emph{compatible} arrangements 
are only those with $n_i>0$ and $s_i=1$ (a detection at location $i$ requires an active detector and at least one photon arriving at it).
If $x_i=0$, only the arrangements with $s_i=1$ and $n_i=0$ or those with $s_i=0$ are compatible (a non-detection may correspond to
an active detector with no photon impinging on it or to an inactive detector).
If $x_i=1$ and $s_i=0$ there are no compatible arrangements (when the detector $i$ is inactive it cannot detect any photons).
The conditions for a  compatible arrangement are summarized in Table \ref{tbl:requirements}.
Furthermore, the equation  $k\equiv\sum_i x_i\leq n$ must
be satisfied: it corresponds to the condition that the numbers of detections cannot be greater than the number of incoming photons.
Then, when $k\leq n$ the probability $\mathcal{P}(\x|n,\s)$ can be written as
\begin{align}
\label{eq_arr_multi}
&\mathcal{P}(\x|n,\s)
=\sum_{\substack{\n\ {\rm compat.}\\{\rm with\ (\x,\s)}}}
\frac{n! }{M^{n} n_1!\cdots n_M!}\delta_{n_1+\ldots+n_M,n}
\\
\notag&=\delta_{|\mathcal I_{10}|,0}\frac{n!}{M^n}
\prod_{i \in \mathcal I_{00}} (\sum_{n_i=0}^n\frac{1}{n_i!})
\prod_{j \in \mathcal I_{11}} (\sum_{n_j=1}^n\frac{1}{n_j!}) \delta_{n_1+\ldots+ n_M,n}
\end{align}
where $\mathcal I_{ab}$ is the set of indices defined by $\mathcal I_{ab}=\{i \mid x_i=a \wedge s_i=b\}$
and $|\mathcal I_{ab}|$ is its cardinality.
The last relation in \eqref{eq_arr_multi} is obtained by explicitly considering  the arrangements $\n$ that are compatible $(\x,\s)$
as reported in table \ref{tbl:requirements}.
An example of the probabilities evaluated for a matrix with $M=9$ and $n=9$ photons is 
reported in the \virgolette{matrix plot} of Fig. \ref{figura_4} (Left), with different colors related to the values of the probabilities.
Rows and columns account for all the possible $2^9$ configurations of $\s$ and $\x$ respectively. 
The fractal structure that is obtained resembles the \emph{Sierpinsky triangle} \cite{sierpinski1915courbe}, where 
the white areas correspond to combinations of $(\x,\s)$ for which no photon arrangement is compatible.

The finding of the largest probability can be simplified by exploiting the symmetries of $\mathcal{P}(\x|n,\s)$.
Indeed, for the values of $\x$ and $\s$ admitting compatible arrangements (namely those values for which
$|\mathcal I_{10}|=0$) 
the probability $\mathcal{P}(\x|n,\s)$ is uniquely determined by the  Hamming weights of $\x$ and $\s$, 
defined as $k=\sum_ix_i$ and $\ell=\sum s_i$ and it is not vanishing only for $k\leq\ell$.
We note that $k$ and $\ell$ correspond to
the numbers of detections and active pixels respectively.
We also define $r$ as the number of inactive pixels, namely $r=M-\ell$.
Since $k$ and $\ell$ range from 0 to $M$,
inequivalent values of $\mathcal{P}(\x|n,\s)$, denoted by $\mathcal{P}(k|n,\ell)$,
can be listed in a reduced $M+1\times M+1$ matrix, simplifying the maximization in \eqref{H(X|n,s)}
from a set with at most $2^M$ elements to a set with at most cardinality $M+1$.

However, the computation of the conditional probabilities remains hard due to the $k+r$ sums 
contained in the r.h.s. of Eq. \eqref{eq_arr_multi}.  We note that $k+r$ may be as large as $M$, 
implying an increasing complexity with the number of detectors.
As we will demonstrate soon, the evaluation of $\mathcal{P}(\x|n,\s)$ can be simplified by the following relation:
\beq
\label{eq_fact_rStirling}
\mathcal{P}(\x|n,\s)=
\left\{\begin{aligned}
&\delta_{|\mathcal I_{10}|,0} \frac{k!}{M^n} {n+r\brace k+r}_r \,,&\quad& {\rm if\ }k\leq n
\\
&0&& {\rm if\ }k> n
\end{aligned}
\right.
\eeq
where 
\begin{align}\label{eq_rStirling}
{n+r\brace k+r}_r=\sum _{j=0}^k  \frac{(-1)^{k-j}}{j!(k-j)!} (r+j)^n\,.
\end{align}
The symbol ${n\brace k}_r$ represents
the \emph{r-restricted Stirling number of the second kind},
defined as the number of partitions of the set $\{1,\cdots, n\}$ into $k$ non-empty disjoint subsets, such that the numbers
$\{1, \cdots , r\}$ are in distinct subsets \cite{brod84dma}. As demonstrated in appendix the $r$-Stirling numbers can be explicitly 
evaluated as the r.h.s. of Eq. \eqref{eq_rStirling}.
We recall that in the equation \eqref{eq_fact_rStirling} we have defined $k=\sum_i x_i$ and $r=M-\sum_is_i$.
With the introduction of the Stirling numbers we therefore achieved a dramatic reduction of the time necessary to compute $\mathcal P(\x|n,\s)$.
Indeed, the evaluation of ${n+r\brace k+r}_r$ in Eq. (\ref{eq_fact_rStirling}) has just a single sum, that must be compared
with the $k+r$ sums required to evaluate Eq. (\ref{eq_arr_multi}).
The rest of the section will be  devoted to demonstrate Eq. \eqref{eq_fact_rStirling}.

We first note that, when $|\mathcal I_{01}|=0$ and $k\leq n$, the number 
$t_{k,r}(n)\equiv \mathcal P(k|n,\ell)\cdot M^n$ corresponds to the number of ways 
of distributing $n$ photons on $|\mathcal I_{00}|+|\mathcal I_{11}|=r+k$ 
pixels such that $k$ of them 
(the active pixels that have $s_j=1$) receive at least one photon. 
Indeed, as indicated by Eq. \eqref{eq_arr_multi} we must consider the events in which the photons arrive only at pixel $D_i$ and $D_j$
with $i \in \mathcal I_{00}$ and $j \in \mathcal I_{11}$, with the extra requirements that at least one photon arrive at each
active pixel $D_j$. 
To evaluate the number $t_{k,r}(n)$ we may proceed as follow.
Let's consider a set with $n$  ``real'' photons and $r$ \virgolette{fictitious} photons.
By definition, the number of partitions of the $n+r$ photons into $k+r$ non-empty disjoint subsets, 
such that the $r$ fictitious photons are always in distinct subsets is given by the 
$r$-Stirling numbers ${n+r\brace k+r}_r$. We note that, due to the presence of \virgolette{fictitious} photons,
$k$ subsets have at least one ``real'' photon, while the remaining $r$ subsets may have any number of ``real'' photons.
Now we can associate a pixel to each of the $k+r$ subsets: in this way, $k$ pixels receive at least one photon,
as required in the definition of $t_{k,r}(n)$. However,
the number ${n+r\brace k+r}_r$ underestimates $t_{k,r}(n)$:
indeed, different associations of the $k$ subsets containing no fictitious photons with the $k$ active pixels will give rise to different 
photon distributions. On the other hand, the $r$ subsets with one fictious photon can be uniquely associated to
the $r$ inactive pixels. Then, $t_{k,r}(n)$ is obtained by multiplying ${n+r\brace k+r}_r$ by the permutation of $k$ pixels, namely 
$t_{k,r}(n)=k!{n+r\brace k+r}_r$. Since $P(k|n,\ell)=t_{k,r}(n)/M^n$, we have demonstrated Eq. \eqref{eq_fact_rStirling}.

\begin{table}[t!]
\centering\begin{footnotesize}
\begin{tabular}{|c|c|}\hline
1& \{2\}, \{3\}, \{4\}, \{$\alpha$,1\}, \{$\beta$\}\\\hline
2& \{2\}, \{3\}, \{4\}, \{$\alpha$\}, \{$\beta$,1\}\\\hline
3& \{1\}, \{3\}, \{4\}, \{$\alpha$,2\}, \{$\beta$\}\\\hline
4& \{1\}, \{3\}, \{4\}, \{$\alpha$\}, \{$\beta$,2\}\\\hline
5& \{1\}, \{2\}, \{4\}, \{$\alpha$,3\}, \{$\beta$\}\\\hline
6& \{1\}, \{2\}, \{4\}, \{$\alpha$\}, \{$\beta$,3\}\\\hline
7&\{1\}, \{2\}, \{3\}, \{$\alpha$,4\}, \{$\beta$\}\\\hline
8&\{1\}, \{2\}, \{3\}, \{$\alpha$\}, \{$\beta$,4\}\\\hline 
9&\{1,4\}, \{2\}, \{3\}, \{$\alpha$\}, \{$\beta$\}\\\hline
10&\{1\}, \{2,4\}, \{3\}, \{$\alpha$\}, \{$\beta$\}\\\hline
11&\{1\}, \{2\}, \{3,4\} , \{$\alpha$\}, \{$\beta$\}\\\hline
12& \{1,2\}, \{3\}, \{4\}, \{$\alpha$\}, \{$\beta$\}\\\hline
13& \{1\}, \{2,3\}, \{4\}, \{$\alpha$\}, \{$\beta$\}\\\hline
14& \{2\}, \{3,1\}, \{4\}, \{$\alpha$\}, \{$\beta$\}\\\hline
\end{tabular}
\caption{In the Table, we cosider the case of $M=6$ pixels, $n=\{1,2,3,4\}$ photons, $\x=\{1,1,1,0,0,0\}$ and $\s=\{1,1,1,1,0,0\}$. 
We list all the possible
 partitions of the set $\{1,2,3,4,\alpha,\beta\}$ in 5 subsets, such that the fictitious photons are in distinct subsets.
}
\end{footnotesize}
\end{table}

To better illustrated the procedure, we show in Table II the case of $M=6$ pixels, 
$n=4$ photons, $\x=\{1,1,1,0,0,0\}$ and $\s=\{1,1,1,1,0,0\}$. 
In this case, $r=2$ and $k=3$. Since $x_4=0$ and $s_4=1$, no photon must impinge on $D_4$. 
To evaluate $t_{k,r}(n)$, we must count all the distributions of 4 photons on 5 detectors, 
such that three of them ($D_1$, $D_2$ and $D_3$) receive at least one photon.
Since $r=2$, two fictitious photons, labelled $\alpha$ and $\beta$, are added to the set of the real photons. 
In Table II we list
all the partitions of the set $\{1,2,3,4,\alpha,\beta\}$ in 5 subsets, such that the fictitious photons are in distinct subsets.
The number of partitions is indeed counted by 
${6 \brace 5}_2=14$. The sets with $\alpha$ and  $\beta$ can be uniquely associated with the inactive pixels $D_5$ and $D_6$. 
Indeed, the Table accounts for all the inequivalent ``real'' photon arrangements on the inactive pixels.
However,  the Table does not take into account
inequivalent ``real'' photon arrangements on the active pixels: for instance,
in partition 1 we have
the three possible associations with the detectors: $\left\{\{2\} \rightarrow D_1, \{3\} \rightarrow D_2, \{4\}\rightarrow D_3\right\}$,
$\left\{\{2\} \rightarrow D_3, \{3\} \rightarrow D_1, \{4\}\rightarrow D_2\right\}$
and $\left\{\{2\} \rightarrow D_2, \{3\} \rightarrow D_3, \{4\}\rightarrow D_1\right\}$; the same happens for the
other listed partitions.
Then, the number of permutations of the $k$ active pixel should be considered, in order to obtain 
$t_{k,r}(n)=k!{n+r\brace k+r}_r=84$.

\section{Discussion}\label{Sec_4}
By using the results presented in the previous section we are now ready to evaluate the quantum randomness of the generator.
By further assuming that the detectors have all the same efficiency (such that $\widetilde\eta_i\equiv \widetilde\eta$), 
the conditional min-entropy simplifies to
\beq
\label{eq_entropy_2}
\begin{aligned}
&H_\text{min}(X|N,S)=-\log_2\left[\sum_{n=0}^{\infty}\mathcal{P}_N(n)\sum_{r=0}^M
\binom{M}{r}\times\right.
\\\
&\left.\quad\quad\widetilde\eta^{M-r}(1-\widetilde\eta)^{r} \max_{k}
\frac{k!}{M^n} {n+r\brace k+r}_r \right] \,,
\end{aligned}
\eeq
that evaluates the amount of bits which can be considered random and secure after the application of a quantum randomness extractor.

\begin{figure}[t!]
\centering\includegraphics[width=\columnwidth]{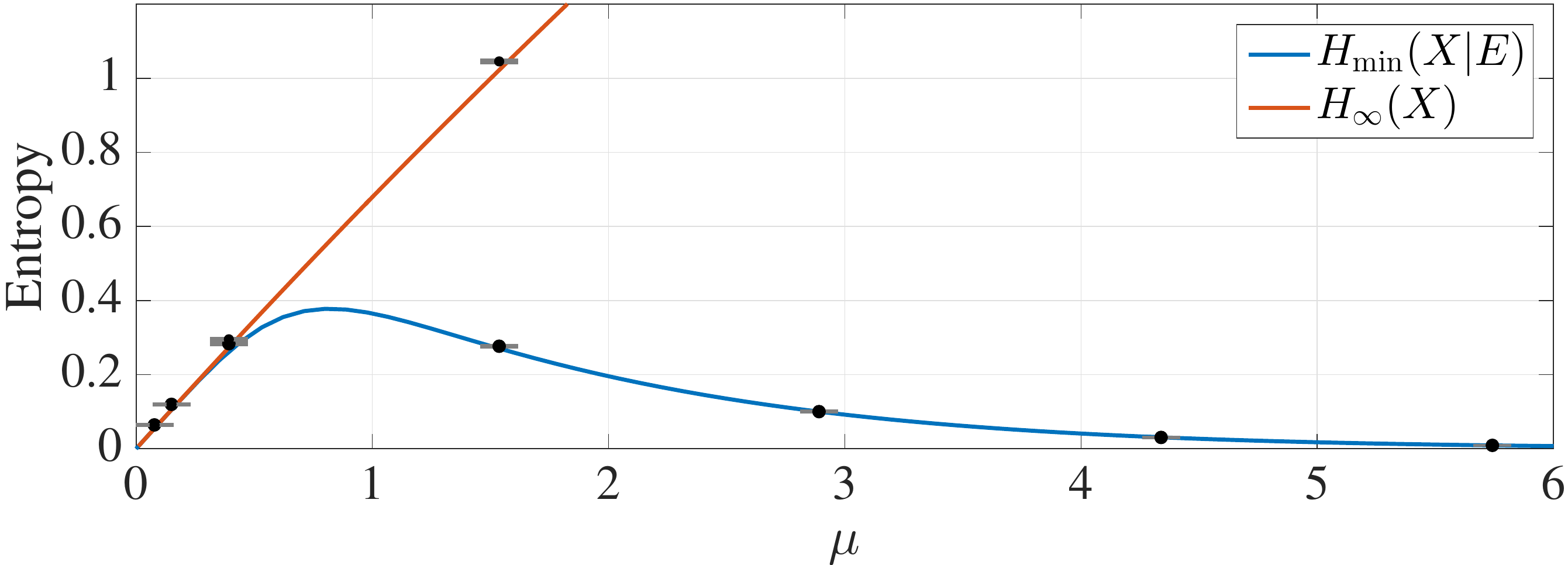}
\caption{The classical min-entropy, $H_{\infty}(X)$ (red line), with the experimental points and the conditional one, 
$H_{\text{min}}(X|C)$ (blue line), are plotted as function of $\mu$, the average number of photons emitted by the source 
per integration time and per pixel. 
The plot considers the sub-matrix of $3 \times 3$ pixels of Fig. \ref{figura_1}.}
\label{figura_5}
\end{figure}

In Fig. \ref{figura_5} we compare the classical min-entropy $H_{\infty}(X)$ (red solid line) and $H_{\text{min}}(X|C)$ (blue solid line) as function of the parameter $\mu$, for the $3 \times 3$ pixel sub-matrix considered in Fig. \ref{figura_1}. 
As shown in Fig.\ref{figura_5}, when the eavesdropper has information on the emitted photon number and can gate the
detectors,
the classical min-entropy is not suitable to estimate the real content of randomness as it yields a too optimistic estimation.
In particular we observe that by using $H_{\infty}(X)$, 
the user might be induced to increase the value of $\mu$ to maximize the entropy. 
In reality, by doing so, the user gives to Eve a dramatic guessing advantage as only $\approx 6\cdot10^{-12}$ bits out of 9 can be considered secure
for $\mu\approx\mu_{\rm opt}$.
In fact, when the mean number of photons per pixel is high, although the input state is pure, 
all the active pixels will output a bit $x_i=1$ with high probability, while 
all the inactive pixels will output a bit $x_i=0$ with certainty: by knowing the detector status $\s$, Eve may
guess with high probability the output string $\x$.

As it can be derived from the plot, the best strategy to obtain a high secure rate it is to keep low the average number of photon per pixel: in this way, although Eve already knows that the inactive pixels will output $x_i=0$, thanks to the purity of the state, 
she cannot predict the outcomes of the active pixels. 
\begin{figure}[h!]
\centering\includegraphics[width=0.5\columnwidth]{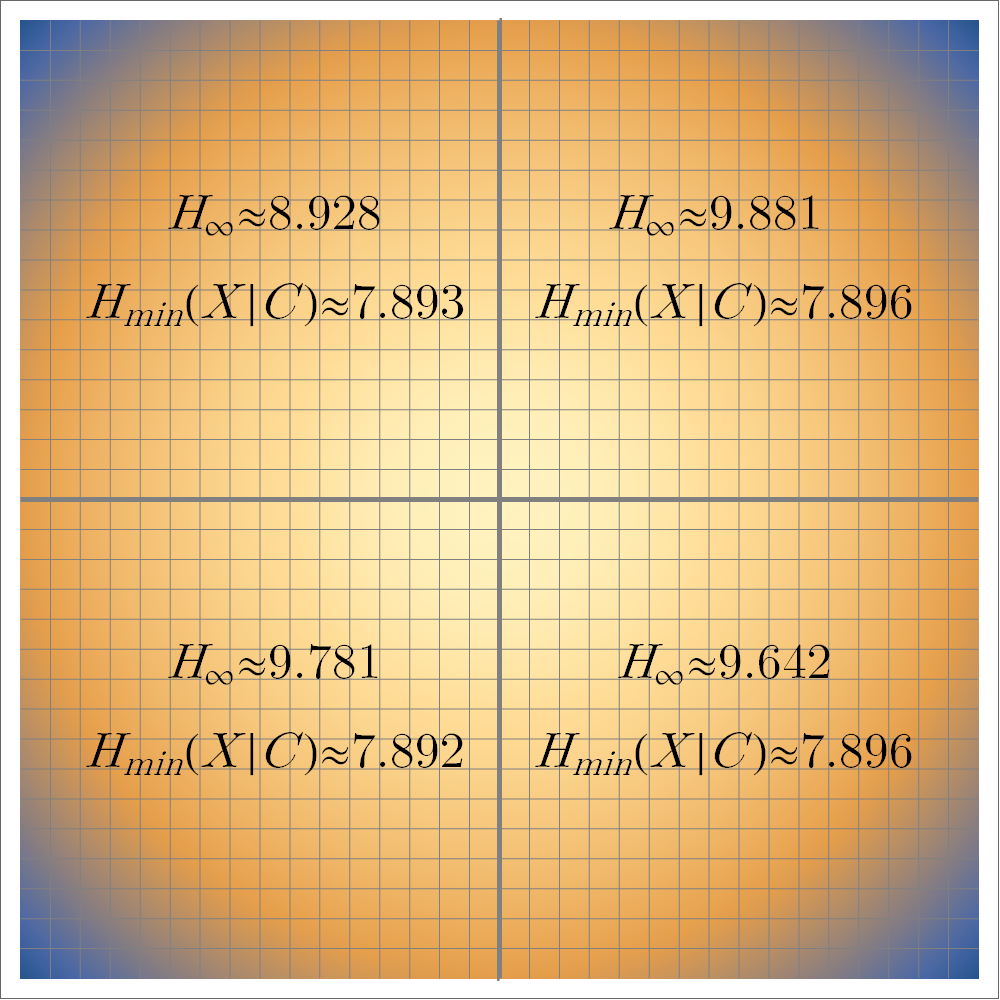}
\caption{Classical and quantum min-entropy evaluated on the four $16 \times 16$ pixels sub-matrices of the Camera.}
\label{figura_6}
\end{figure}
In this condition, we applied the efficient relation of Eq. (\ref{eq_entropy_2}) 
on a large number of detectors and we estimated the entropy of the four $16 \times 16$ pixels sub-matrices 
as if they were illuminated by an uniform intensity
(an improved code for evaluating Eq. \eqref{eq_entropy_2} is required for a larger number of pixels). 
The comparison of the classical and quantum min-entropies is presented
in Fig. \ref{figura_6}. 
We conservatively assumed that the whole sensor was illuminated with the largest value of $\mu$ registered: in this way it is possible to increase the relative secure rate.
By this approach we were able to generate secure random numbers at a rate of $R\simeq 1.547~Mbit/s$.
We note that such rate can be improved by using a laser light at a different wavelength, where the detection efficiency is higher.
Indeed, when the measured detection efficiency is high, Eve must keep the detector active with high probability, giving her less chance to
guess the output string $\x$.

\section{Conclusions}
We here presented  a protocol which enable the user of a multi-detector QRNG to extract secure random bits  for cryptographic applications. 
In our framework an eavesdropper may have access to the \virgolette{classical information}  of the QRNG,
that can be used to enhance his chances to guess the generator outcomes. 
Such \virgolette{classical information} is represented by the number of photons emitted by the source and the detector status.
Indeed, in our model, the measured detector inefficiencies are modeled by perfect detectors that are randomly activated by the adversary.
As we demonstrated, an efficient combinatorial method can be used to calculate the 
conditional min-entropy. Once the quantum min-entropy is known, the generated sequence must be post-processed by
a randomness extractor that will generate bits with the rate imposed by $H(X|C)$.
In this ``paranoid scenario'', we were able to generate secure random numbers at a rate of $R\simeq 1.547~Mbit/s$.
It is worth noticing that, if the information available to Eve is reduced (for instance by assuming that
she does not known the number of photons emitted by the source or by assuming that she cannot control
the detector gating), the secure key rate will be increased.

 By comparing the conditional min-entropy $H(X|C)$ and the commonly used classical min-entropy $H_\infty(X)$, 
we have shown how the use of the latter, while it may increase the generation rate up to 
$R_{\rm gen}= 50.176 M bit/s$, it might endanger the QRNG security. 

\section{Aknowlegements}
We thank Micro Photon Devices for the loan of the SPC$^2$-camera.

\appendix
\section{Evaluation of the r-restricted Stirling numbers}
As reported in Eq. (31) of \cite{brod84dma},
the $r$-restricted Stirling numbers of the second kind satisfy the following relation
\beq
\label{s1}
{n \brace k}_r=\sum_{i=0}^{n-r}
\binom{n-r}{i}{n-p-i \brace k-p}_{r-p}p^i\,,
\eeq
for any integer $p$ such that $0\leq p\leq r$.
By replacing $n\rightarrow n+r$, $k\rightarrow k+r$ and using $p=r$ we obtain
${n+r \brace k+r}_r=\sum_{i=0}^{n}
\binom{n}{i}{n-i \brace k} r^i$.
Now, we may exploit the property that the Stirling numbers ${n \brace k}$ can be evaluated by the explicit formula~\cite{shar68jct}
\beq
{n \brace k} =\frac{1}{k!}\sum_{j=0}^k(-1)^{k-j}\binom{k}{j}j^n\,.
\eeq
Then \eqref{s1} may be written as
\beq
\begin{aligned}
{n+r \brace k+r}_r=
&\frac{1}{k!}\sum_{j=0}^k(-1)^{k-j}
\binom{k}{j}
\sum_{i=0}^{n}\binom{n}{i}
j^{n-i}r^i\,
\\
=&\frac{1}{k!}\sum_{j=0}^k(-1)^{k-j}
\binom{k}{j}
(j+r)^n
\\
\end{aligned}
\eeq
We have thus demonstrated Eq. \eqref{eq_rStirling} of the main text.

\end{document}